\documentclass[
  ,final 
]{aipproc}
\layoutstyle{8x11single}
\usepackage{graphicx}
\usepackage{bm}
\usepackage{amsmath}
\usepackage{amssymb}
\usepackage{color}
\begin{document}

\title{Slow Dynamics Near Jamming}


\author{Kuniyasu Saitoh}
{address={Faculty of Engineering Technology, University of Twente, Enschede, the Netherlands}}
\author{Vanessa Magnanimo}
{address={Faculty of Engineering Technology, University of Twente, Enschede, the Netherlands}}
\author{Stefan Luding}
{address={Faculty of Engineering Technology, University of Twente, Enschede, the Netherlands}}

\begin{abstract}
Static and dynamic properties of two-dimensional bidisperse dissipative particles
are numerically studied near the jamming transition. We investigate the dependency
of the critical scaling on the ratio of the different diameters and find a new scaling
of the maximum overlap (not consistent with the scaling of the mean overlap). The
ratio of kinetic and potential energies drastically slows down near the jamming transition,
i.e. the relaxation time diverges at the jamming point.
\end{abstract}
\maketitle

\section{Introduction}

Jamming is an universal feature of both thermal and athermal systems, for instance,
glasses, granular particles, emulsions, colloidal suspensions, foams, etc. where
constituents are arrested in disordered states so that the material obtains rigidity.
The jamming transition is governed by temperature, density and external loads, and
a lot of systems can be mapped onto a unified phase diagram \cite{ph0,ph1,ph2}.

Jamming of athermal systems, i.e. granular particles \cite{gn0,gn1,gn2,gn3},
emulsions \cite{co0,co1} and foams \cite{fo0,fo1}, occurs at zero temperature at
the critical density (area fraction) $\phi_c$ \cite{ph0}. At this point, each particle begins to touch
its neighbors and the mean coordination number $z$, defined as the average number
of contacts per particle, jumps from zero to the isostatic value $z_c=2d$ in $d$-dimensions
\cite{am0}. Some macroscopic variables also indicate the acquisition of rigidity,
for example, the pressure $p$ and the shear modulus $G$ start to increase from zero, and the
bulk modulus $K$ discontinuously jumps to a non-zero value \cite{gn0,gn1,gn2,gn3}.
Above this threshold, the excess coordination number $\Delta z=z-z_c$, $p$, $G$ and $K$ respectively
scale as $\Delta z\sim\Delta\phi^{1/2}$, $p\sim\Delta\phi^\psi$, $G\sim\Delta\phi^\gamma$
and $K\sim\Delta\phi^\lambda$ with the distance from the jamming point $\Delta\phi=\phi-\phi_c$.
In the case of monodisperse particles, the first
peak of the radial distribution function $g_1$ scales as $g_1\sim\Delta\phi^{-1}$
\cite{gr0,gr1,gr2,gr3}. In the case of bidisperse particles, one can also see a
similar divergence of the radial distribution function with the scaled distance
\cite{gn3}. Moreover, the mean overlap between particles $\langle\delta\rangle$
linearly scales as $\langle\delta\rangle\sim\Delta\phi$.

Those power law scalings are analogous to those found in critical phenomena. However, some
variables show discontinuous changes at the critical point as in first-order phase
transitions. Furthermore, the critical exponents $\psi$, $\gamma$ and $\lambda$ depend on the
interparticle forces, which suggests that the jamming transition is entirely different from
usual critical phenomena \cite{gn0,gn1,gn2,gn3}.

Even though the critical scalings above the jamming transition are well established,
not much attention has been paid to the critical amplitudes. Many previous works on
bidisperse systems only focused on the particular case that the ratio of the different diameters
equals $\rho=1.4$. Furthermore, the dynamic properties are not understood yet.

In this paper, we study the static and dynamic properties of two-dimensional
bidisperse particle systems by numerical simulations and investigate the dependency of the
critical amplitudes on $\rho$. We also study the dynamic properties near the jamming
transition, where we adopt the ratio of kinetic to potential energies to quantify relaxation
and show its drastic slowing down near the jamming point.
\section{Molecular Dynamics Simulation}
We investigate two-dimensional packings of bidisperse dissipative particles using molecular
dynamics (MD) simulations. Our strategy is to increase the diameter of each particle until a
desired area fraction $\phi$ is obtained. After the desired area fraction is obtained, the
kinetic energy decreases to zero due to the dissipative forces between particles and the
system relaxes to its static state. We study dynamic properties during the relaxation and
static properties after the relaxation.

At first, we prepare a binary mixture of large and small particles with initial diameters
$\sigma_\mathrm{L}(0)$ and $\sigma_\mathrm{S}(0)$, respectively, where the total number of
particles is $N=32768$. We randomly distribute them in a $L\times L$ periodic box one-by-one,
avoiding overlap between particles. Then, we slowly increase the diameter of each particle
\cite{lsa,hd0,hd1,luding4} as:
\begin{equation}
\dot{\sigma}_a(t)=g_r\sigma_a(0)\hspace{0.5cm}(a=\mathrm{L}~\mathrm{or}~\mathrm{S})
\label{eq:lgr}
\end{equation}
with a constant growth rate $g_r$. Because we fix the mass density of particles, the mass
also increases while we increase the diameter $\sigma_a(t)=\sigma_a(0)(g_rt+1)$.
Therefore, the size ratio
$\rho\equiv\sigma_\mathrm{L}(t)/\sigma_\mathrm{S}(t)=\sigma_\mathrm{L}(0)/\sigma_\mathrm{S}(0)$
does not change throughout our simulations.

Each particle $i$ can be in contact with several other particles $j$ and moves according to the equation of motion
\begin{equation}
m_i(t) \ddot{\mathbf{x}}_i = \sum_j \left\{ k_n \delta_{ij} - \eta_n (\mathbf{v}_{ij}
\cdot \mathbf{n}_{ij}) \right\} \mathbf{n}_{ij}~,\label{eq:normal_eqmotion}
\end{equation}
where $m_i(t)$, $k_n$, $\eta_n$, $\delta_{ij}$,
$\mathbf{n}_{ij}=(\mathbf{x}_i-\mathbf{x}_j)/|\mathbf{x}_i-\mathbf{x}_j|$,
and $\mathbf{v}_{ij}=\dot{\mathbf{x}}_i-\dot{\mathbf{x}}_j$
are the mass of particle $i$, the spring constant,
the viscosity coefficient, the overlap,
the normal unit vector, and the relative velocity, respectively.
The vector $\mathbf{x}_i$ is the position of particle $i$,
and each dot above $\mathbf{x}$ represents the time derivative.

When the area fraction reaches the desired value $\phi$ at $t=t_0$, we stop to increase the
diameter of each particle. Then, the kinetic energy of the system decreases due to the
dissipative forces and we assume the system is relaxed to its static state
if the ratio of kinetic to potential energies becomes lower than $10^{-6}$. In the following,
we scale mass, length and time by $m_u=m_\mathrm{L}(t_0)$, $l_u=\sigma_\mathrm{L}(t_0)$ and
$t_u=m_\mathrm{L}(t_0)/\eta_n$, respectively, and use $g_r=10^{-4}/t_u$,
$k_n=4.0\times 10^{5} m_u/t_u^2$ and $\eta_n=m_u/t_u$.

From Eq. (\ref{eq:normal_eqmotion}), the restitution coefficient in the relaxation stage
($t>t_0$) defined as the ratio of speeds after and before a collinear collision is calculated as \cite{dem}
\begin{equation}
e_n = e^{-\eta_0t_c}~,\label{eq:e}
\end{equation}
where $t_c=\pi/\sqrt{(k_n/m_{ij})-\eta_0^2}$ and $\eta_0=\eta_n/(2m_{ij})$ are a typical
response time and the rescaled viscosity coefficient with the reduced mass
$m_{ij}=m_i(t_0)m_j(t_0)/(m_i(t_0)+m_j(t_0))$. Although the restitution
coefficients between two small particles, two large particles, and small and large particles
are slightly different, we find $e_n=0.99\pm0.003$ in the three cases since dissipation
is rather weak.

\section{Critical scaling}
Changing the values of $\phi$ and $\rho$ between $0.8\le\phi\le0.9$ and $1.2\le\rho\le2.4$,
respectively, we repeat our simulations and measure the mean overlap $\langle\delta\rangle$,
the mean coordination number $z$, the pressure $p$, the first peak of the radial distribution
function with the scaled distance $g_+$ and the maximum overlap $\delta_\mathrm{m}$, after each
system is relaxed to its static state. We define the critical area fraction $\phi_c$ as the
point at which $\langle\delta\rangle$ starts to increase from zero and $z$ jumps from zero
to the isostatic value $z_c=4$. It should be noted that we remove \emph{rattlers} of which
the number of contacts are less than $3$ from the statistics, because they do not contribute
to the contact- and force-chain networks \cite{gn0,gn1,gn2,gn3}.

\begin{figure}
\includegraphics[width=8cm]{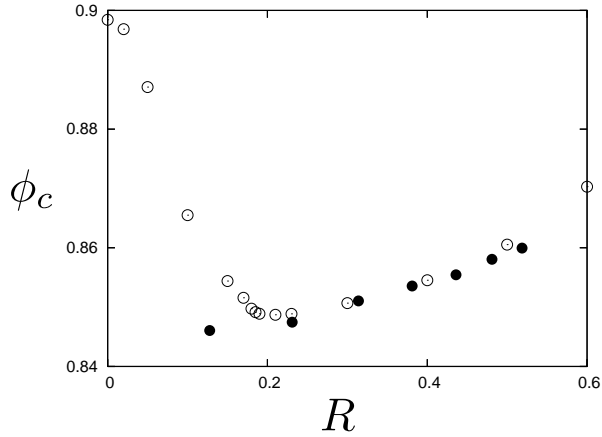}%
\caption{Critical area fraction $\phi_c$ as a function of $R$, where the closed circles
are the results of our simulations and the open circles are the results of another
simulation of bidisperse hard spheres in two dimensions \cite{numax}. \label{fig:phic}}
\end{figure}

The critical area fraction depends on both $\rho$ and $g_r$.
Figure \ref{fig:phic} shows $\phi_c$ as a function of the relative standard deviation
\begin{equation}
R\equiv\frac{\sqrt{\langle\sigma^2\rangle-\langle\sigma\rangle^2}}{\langle\sigma\rangle}~,
\label{eq:R}
\end{equation}
where $\langle\sigma\rangle=(1+\rho)\sigma_\mathrm{S}(t_0)/2$ and $\langle\sigma^2\rangle
=(1+\rho^2)\sigma_\mathrm{S}(t_0)^2/2$ are the mean diameter and the
mean square of diameter, respectively. In this figure, the closed circles are our results
and the open circles are the results of another simulation of bidisperse hard spheres in
two dimensions \cite{numax}, where a large discrepancy between these results can be seen
below $R=0.2$, mainly caused by the dependency of $\phi_c$ on $g_r$. In this paper,
we do not discuss the rate dependency of $\phi_c$, however, it is known that $\phi_c$ strongly
depends on $g_r$, if $R$ is small (see Ref. \cite{luding2} for a more detailed discussion).
We also list our results of $\phi_c$ for different $\rho$ and $R$ in Table \ref{tab:ampl}.

Above $\phi_c$, $\langle\delta\rangle$ scales as \cite{gn0,gn1,gn2,gn3}
\begin{equation}
\langle\delta\rangle = B_\delta(\rho) \Delta\phi^\mu~, \label{eq:scaling_d}
\end{equation}
where $\mu$ and $B_\delta(\rho)$ are the \emph{critical exponent} and the \emph{critical
amplitude}, respectively (see Table \ref{tab:ampl}). From our simulations, we find $\mu\simeq 1.0$,
i.e. $\langle\delta\rangle$ depends linearly on $\Delta\phi$ (see Table \ref{tab:powers}).

\begin{figure}
\includegraphics[width=17cm]{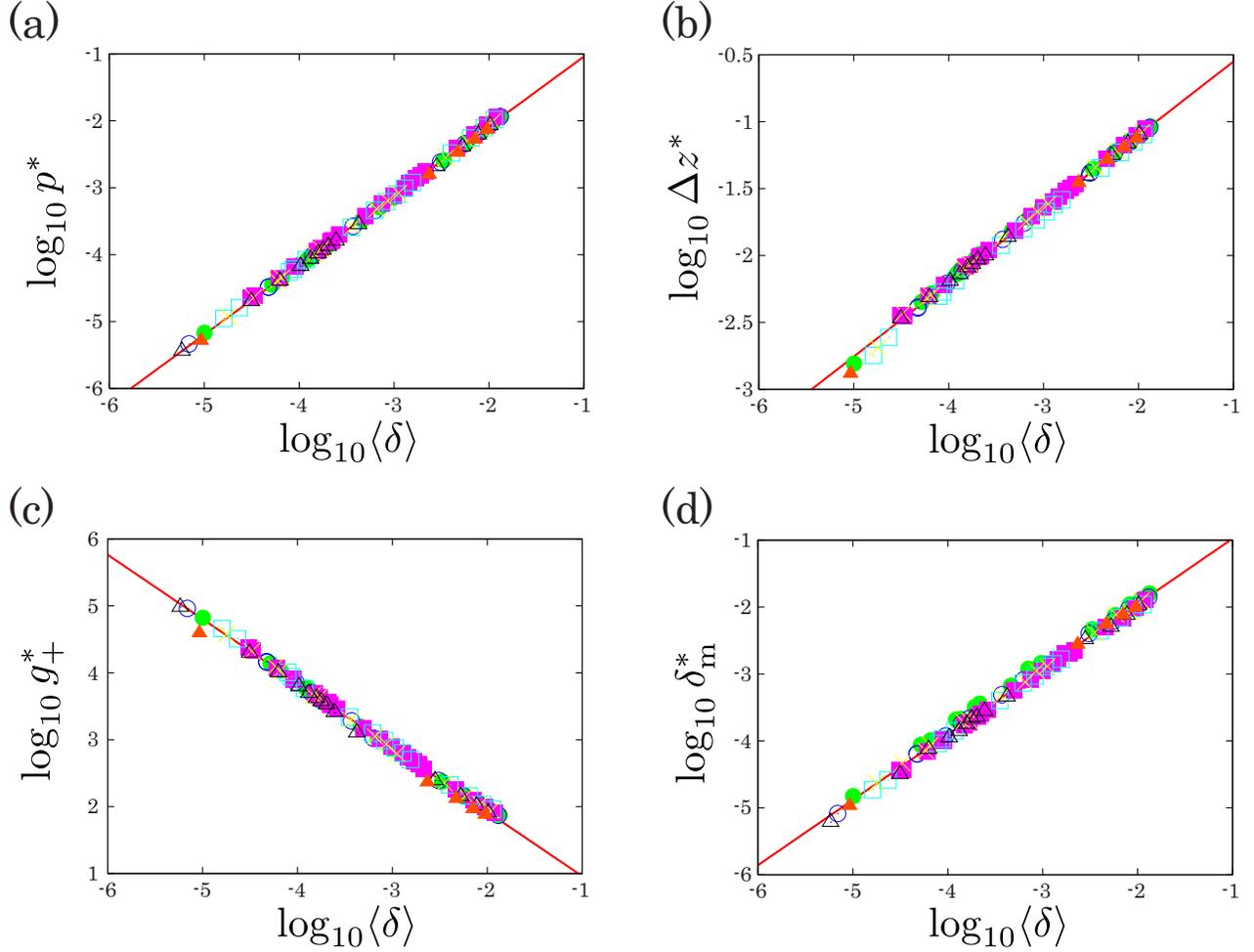}
\caption{(Color online) Double logarithmic plots of (a) $p^\ast$, (b) $z^\ast$,
(c) $g_+^{\ast}$ and (d) $\delta_\mathrm{m}^\ast$ as functions of $\langle\delta\rangle$.
The solid circles, open circles, solid squares, open squares,
crosses, open triangles and solid triangles are the data for $\rho=1.2$, $1.4$,
$1.6$, $1.8$, $2.0$, $2.2$ and $2.4$, respectively. \label{fig:power}}
\end{figure}

We define the pressure as the virial pressure $p=\sum_{i<j}r_{ij}f_{ij}/L^2$
with the interparticle distance $r_{ij}$ and force $f_{ij}$ \cite{comliq} and introduce
the excess coordination number as $\Delta z\equiv z-z_c$. Because we use
bidisperse particles, the usual radial distribution function has three peaks around $\sigma_\mathrm{S}$,
$\sigma_\mathrm{L}$ and $(\sigma_\mathrm{L}+\sigma_\mathrm{S})/2$, respectively. However,
if we introduce the scaled distance between particles $i$ and $j$ as $\xi=r_{ij}/(r_i+r_j)$,
where $r_i$ and $r_j$ are the radii of particles $i$ and $j$, respectively, the radial
distribution function of $\xi$ has the first peak $g_+$ around $\xi=1$.
Then, we find $p$, $\Delta z$ and $g_+$ scale as
\begin{eqnarray}
p &=& A_p(\rho) \langle\delta\rangle^\psi~, \label{eq:scaling_p}\\
\Delta z &=& A_z(\rho) \langle\delta\rangle^\zeta~, \label{eq:scaling_z}\\
g_+ &=& A_+(\rho)\langle\delta\rangle^{-\eta_+}~, \label{eq:scaling_gp}
\end{eqnarray}
respectively, where $A_p(\rho)$, $A_z(\rho)$ and $A_+(\rho)$ are the critical amplitudes
(see Table \ref{tab:ampl}), and $\psi\simeq 1$, $\zeta\simeq 1/2$ and $\eta_+\simeq 1$ are
the critical exponents (see Table \ref{tab:powers}). Figures \ref{fig:power} (a)-(c) display
$p^\ast \equiv p/A_p(\rho)$, $\Delta z^\ast\equiv\Delta z/A_z(\rho)$ and $g_+^{\ast}\equiv g_+/A_+(\rho)$,
respectively. Since $\langle\delta\rangle\sim\Delta\phi$, these results confirm $p\sim\Delta\phi$,
$\Delta z\sim\Delta\phi^{1/2}$, $g_+\sim\Delta\phi^{-1}$ and
$g_+\langle\delta\rangle\approx\mathrm{const}$ \cite{gn0,gn1,gn2,gn3}.

From our simulations, we also find $\delta_\mathrm{m}$ scales as
\begin{equation}
\delta_\mathrm{m} = A_\mathrm{m}(\rho)\langle\delta\rangle^\lambda \label{eq:scaling_x}
\end{equation}
with the critical amplitude $A_\mathrm{m}(\rho)$ (see Table \ref{tab:ampl}).
Figure \ref{fig:power} (d) shows $\delta_\mathrm{m}^\ast\equiv\delta_\mathrm{m}/A_\mathrm{m}(\rho)$,
where we find $\lambda\simeq 1$ similar to $\mu$ (see Table \ref{tab:powers}).
The ratios $A_\mathrm{m}(\rho)/B_\delta(\rho)$ are not constant (see Table \ref{tab:ampl}),
which means that the probability distribution functions of the overlap are not self-similar,
but change shape with increasing $\rho$ \cite{luding5}.
\begin{table}
\begin{tabular}{ccccccccc} \hline
$\rho$ & $R$ & $\phi_c$ & $B_\delta$ & $A_p/k_n$ & $A_z$ & $A_+$ & $A_\mathrm{m}$ & $A_\mathrm{m}/B_\delta$ \\ \hline
$1.2$ & $0.127$ & $0.846054$ & $0.262691$ & $0.062944$ & $9.623507$ & $0.108473$ & $3.271311$ & $12.453076$ \\
$1.4$ & $0.230$ & $0.847454$ & $0.247794$ & $0.062175$ & $9.789582$ & $0.112457$ & $3.839339$ & $15.494075$ \\
$1.6$ & $0.313$ & $0.851053$ & $0.234816$ & $0.058778$ & $9.696656$ & $0.107054$ & $4.508161$ & $19.198696$ \\
$1.8$ & $0.381$ & $0.853553$ & $0.240875$ & $0.061023$ & $10.680258$& $0.102261$ & $4.554234$ & $18.907043$ \\
$2.0$ & $0.435$ & $0.855449$ & $0.246302$ & $0.062333$ & $9.661944$ & $0.109497$ & $4.378124$ & $17.775430$ \\
$2.2$ & $0.481$ & $0.858054$ & $0.278428$ & $0.067636$ & $9.986629$ & $0.122986$ & $5.144424$ & $18.476676$ \\
$2.4$ & $0.518$ & $0.859950$ & $0.268369$ & $0.073998$ & $10.649745$& $0.207261$ & $5.304872$ & $19.767081$ \\
\hline
\end{tabular}
\caption{The critical area fraction and the critical amplitudes.}
\label{tab:ampl}
\end{table}
\begin{table}
\begin{tabular}{cccccc} \hline
function & base & exponent    & value & deviation & error $[\%]$ \\ \hline
$\langle\delta\rangle$ & $\Delta\phi$           & $\mu$    & $1.006$ & $\pm 0.003$ & $0.349$ \\
$p$                    & $\langle\delta\rangle$ & $\psi$   & $1.039$ & $\pm 0.001$ & $0.065$ \\
$\Delta z$             & $\langle\delta\rangle$ & $\zeta$  & $0.551$ & $\pm 0.001$ & $0.124$ \\
$g_+$                  & $\langle\delta\rangle$ & $\eta_+$ & $0.960$ & $\pm 0.001$ & $0.148$ \\ 
$\delta_\mathrm{m}$    & $\langle\delta\rangle$ & $\lambda$& $0.976$ & $\pm 0.001$ & $0.164$ \\
\hline
\end{tabular}
\caption{Estimated exponents and errors.}
\label{tab:powers}
\end{table}
\section{Critical slowing down near jamming}
After we stop to increase the diameter of each particle at $t=t_0$, the system
relaxes to its static state. To quantify the relaxation dynamics, we introduce
a dimensionless energy
\begin{equation}
\chi(t)\equiv \frac{K(t)/U(t)}{K(t_0)/U(t_0)}~,\label{eq:def_phit}
\end{equation}
where $K(t)$ and $U(t)$ are the kinetic and potential energies, respectively.
The ratio $K(t)/U(t)$ can be used to estimate $\phi_c$, because $U(t)$
drops to zero and $K(t)/U(t)$ diverges if $\phi<\phi_c$ \cite{luding3}. The
function $\chi(t)$ is the ratio $K(t)/U(t)$ scaled by the value at $t_0$. In
the following, we redefine $t_0=0$ for simplicity and study the case of $\rho=1.4$.

\begin{figure}
\includegraphics[width=17cm]{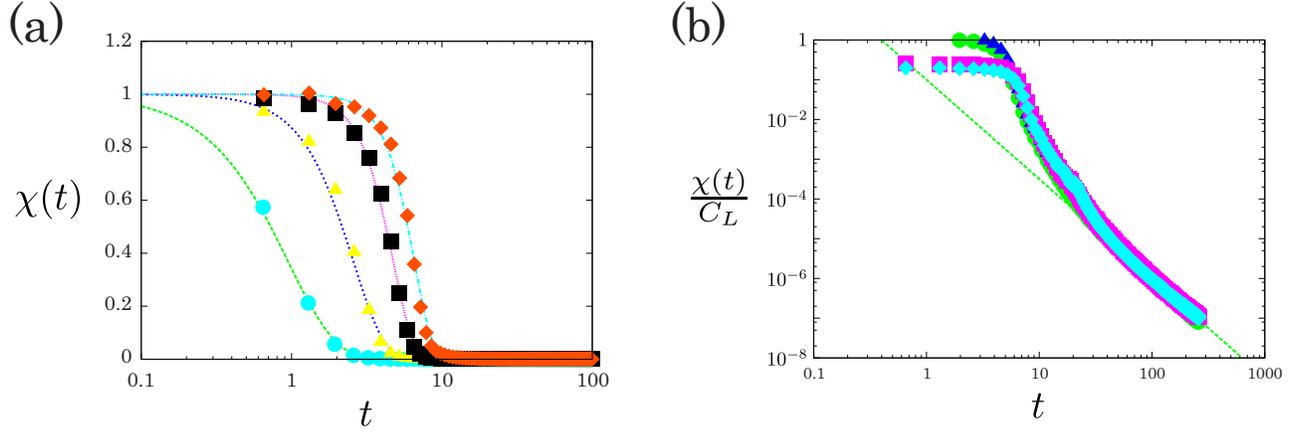}%
\caption{(Color online) (a) $\chi(t)$ against logarithmic time above $\phi_c$,
where the circles, triangles, squares and diamonds are the results of simulations
with $\Delta\phi=1.2\times10^{-2}$, $2.5\times10^{-3}$, $4.4\times10^{-4}$ and
$1.4\times10^{-4}$, respectively. The dotted lines represent Eq. (\ref{eq:stretched}).
(b) Double logarithmic plot of $\chi(t)/C_L$, where the circles, triangles, squares
and diamonds are the results of simulations with $\Delta\phi=5.5\times 10^{-4}$,
$4.5\times10^{-4}$, $2.5\times10^{-4}$ and $1.5\times10^{-4}$, respectively. The
dotted line represents the power law decay with the exponent~$2.541$.
Here, we used $\rho=1.4$. \label{fig:early}}
\end{figure}

\begin{figure}
\includegraphics[width=16cm]{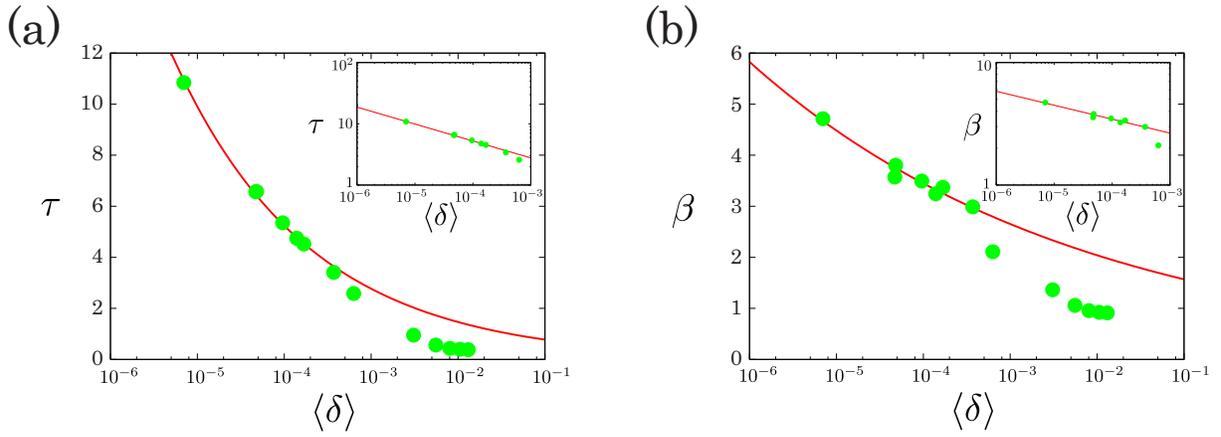}%
\caption{(Color online) (a) $\tau$ and (b) $\beta$ against logarithmic $\langle\delta\rangle$,
where the insets show the double logarithmic plots of them.
The solid circles are our simulation results and the solid lines represent
$\tau\propto\langle\delta\rangle^{-\kappa}$ and $\beta\propto\langle\delta\rangle^{-\nu}$
with $\kappa=0.274\pm 0.006$ and $\nu=0.112\pm 0.008$, respectively. Here, we used
$\rho=1.4$. \label{fig:taub}}
\end{figure}

Figure \ref{fig:early}(a) shows our results of $\chi(t)$ against logarithmic time,
where the decrease of $\chi(t)$ drastically slows down near the jamming point.
To quantify the slowing down, we fit $\chi(t)$ by a stretched exponential function
\begin{equation}
\chi(t) = e^{-(t/\tau)^\beta}~, \label{eq:stretched}
\end{equation}
where $\tau$ and $\beta$ are the relaxation time and the stretching exponent,
respectively. Figure \ref{fig:taub} displays both quantities as functions of 
$\langle\delta\rangle$.
Slightly above $\phi_c$ ($0<\Delta\phi<10^{-3}$), both
$\tau$ and $\beta$ scale as $\tau\propto\langle\delta\rangle^{-\kappa}$
and $\beta\propto\langle\delta\rangle^{-\nu}$ with $\kappa\simeq 0.27$ and
$\nu\simeq 0.11$, respectively.
However, given the rather narrow range of $\tau$ and $\beta$, power laws
with such small exponents are not of significance and require a wider range before
conclusions can be drawn.
It should be noted that $\beta>1$ near the jamming
point and $\chi(t)$ decays faster than exponential with time.

Figure \ref{fig:early}(b) is a double logarithmic plot of the long term asymptotic
behavior of $\chi(t)$ near the jamming point, where we find a power law decay
\begin{equation}
\chi(t) = C_L t^{-\alpha}~. \label{eq:lttail}
\end{equation}
The exponent $\alpha\simeq 2.5$ is not significantly changed by changing $\phi$ and faster
than the decay rate of the kinetic energy in the homogeneous cooling state of granular gases,
where the exponent is given by $2$ \cite{luding1}.

\section{Summary}
In summary, we numerically investigated the static and dynamic properties of
two-dimensional bidisperse particles near the jamming transition and systematically
studied the dependency of the critical exponents and amplitudes on the size ratio
$\rho$. For different size ratios, 
we found different scaling prefactors of the average and maximum overlaps
confirming their different behaviors and thus indicating 
different shapes of the overlap probability
distribution functions \cite{luding5}.

Concerning the dynamics, we used the energy ratio $\chi(t)$ to quantify the 
relaxation of kinetic energy and report that it resembles a stretched exponential
and drastically slows down near the jamming point. 
The asymptotic behavior of $\chi(t)$ resembles a power law decay with
exponent $2.5$, which is faster than the decay rate of the kinetic energy in the
homogeneous cooling state of granular gases.

The cases of monodisperse and polydisperse particles are left to future
study, as is the case for three dimensions. Although the number of particles
in our simulations may be large enough to study the critical area fraction and
the critical scaling, we have to investigate the influence of the system size, the
growth rate and the restitution coefficient elsewhere.
\begin{theacknowledgments}
We thank M. van Hecke, B. Tighe, H. Hayakawa, T. Hatano, N. V. Brilliantov and K. Yazdchi
for fruitful discussions and T. Weinhart for his critical reading.
This work was financially supported by the NWO-STW VICI grant 10828.
\end{theacknowledgments}
%

\bibliographystyle{aipproc}   
\bibliography{rgd28_jamming}

\IfFileExists{\jobname.bbl}{}
 {\typeout{}
  \typeout{******************************************}
  \typeout{** Please run "bibtex \jobname" to optain}
  \typeout{** the bibliography and then re-run LaTeX}
  \typeout{** twice to fix the references!}
  \typeout{******************************************}
  \typeout{}
 }

\end{document}